# Interactive X-Ray and Proton Therapy Training and Simulation

Felix G. Hamza-Lup, Shane Farrar, Erik Leon
*Computer Science and Information Technology,
Armstrong State University, Savannah, GA, USA*
*Felix.Hamza-Lup@armstrong.edu*

**Abstract**

**Purpose:** External beam X-Ray Therapy (XRT) and Proton Therapy (PT) are effective and widely accepted forms of treatment for many types of cancer. However, the procedures require extensive computerized planning. Current planning systems for both XRT and PT have insufficient visual aid to combine real patient data with the treatment device geometry to account for unforeseen collisions among system components and the patient.
**Methods:** The 3D Boundary representation (B-rep) is a widely used scheme to create 3D models of physical objects. 3D B-reps have been successfully used in CAD/CAM and, in conjunction with texture mapping, in the modern gaming industry to customize avatars and improve the gaming realism and sense of presence. We are proposing a cost-effective method to extract patient-specific B-reps in real time and combine them with the treatment system geometry to provide a comprehensive simulation of the XRT/PT treatment room.
**Results**: The X3D standard is used to implement and deploy the simulator on the web, enabling its use not only for remote specialists' collaboration, simulation, and training, but also for patient education.
**Conclusions**: An objective assessment of the accuracy of the B-reps obtained proves the potential of the simulator for clinical use.

## 1. Introduction

Radiation therapy is an effective and widely accepted form of treatment for many types of cancer. External beam radiation therapy (EBRT) [7] is the careful use of radiation from external sources, where X-Rays (called X-Ray Therapy - XRT), or Protons (called Proton Therapy - PT) are administered to the tumors. Recently, the use of XRT has slowly given way to PT for the treatment of certain cancers, such as spinal and eye cancer. The advantage of PT is that higher doses of radiation can be used to destroy and manage malignant tissue with significantly less damage to healthy tissue and vital organs as illustrated in Figure 1, PT has no exit dose since the radiation charge is placed at the tumor location, while XRT has an input and an exit dose, hence the potential for healthy tissue damage is much higher.

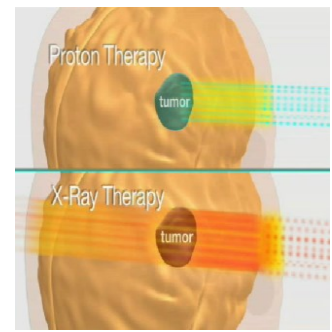

Figure 1. No exit dose for Proton Therapy

Unfortunately, treatment-planning systems for both XRT and PT have limited or no visual aid that combines patient body shape data (i.e. boundary representation) with the treatment system view to provide a detailed understanding of potential collisions and beam angles before the actual therapy begins. While partial patient-specific computed tomography (CT) is available, in general, a full body CT scan is not possible due to the additional radiation exposure to the patient. Therefore, treatment planners often find it difficult to determine precise treatment equipment setup parameters and, in some cases, patient treatment is delayed or postponed due to unforeseen collisions among the system's components or with the patient. Furthermore, the demand for better cancer targeting has created specific immobilization and on-board imaging devices, which can become additional collision sources.

In computer graphics, boundary representation (B-rep) delineates objects in terms of their "skin," that is, the frontier between the model and the environment. B-rep techniques are extensively used in CAD/CAM modeling of solid objects among other methods like cell decomposition, set theoretic and general sweeping [25]. B-reps can be useful in medical simulation for planning and training, more specifically for procedures that involve complex shapes and systems.

We developed and deployed a Web-based XRT simulation system a few years ago [9], and now propose two important upgrades to the system: (1) the introduction of a cost effective, real-time 3D digitization solution for embedding patient B-rep in the simulator and, (2) the addition of a Proton Therapy module in the interactive medical planning and training system.

The choice for implementation of the 3D user interface was eXtensible 3D (X3D) [29], a royalty-free open standard format and web-enabled architecture that provides a system for the storage, retrieval, and playback of real-time 3D graphics content embedded in applications. X3D has a rich set of componentized features that can be tailored for use in engineering and scientific visualization, CAD and architecture, medical visualization, training, simulation, multimedia, entertainment, education, and more. Our proposed solution generates an optimized X3D patient specific B-rep and allows its integration in the simulation system in real time. The paper is structured as follows. Section 2 presents the simulator's scope and recent work by other research groups. Section 3 illustrates the simulator's graphical user interface and accuracy assessment results. Section 4 presents the X3D patient B-rep generation, including patient data acquisition and optimization. Section 5 presents a set of conclusions related to XRT and PT simulator use.

## 2. Rationale and Related Work

External beam radiation therapy planning systems (X-Ray or Proton) make use of 3D visualization of volumetric data. This data usually describes radiation doses that have to be delivered at specific locations to destroy cancerous tissue. Since the treatment's success depends on the accuracy of the planning and radiation delivery, physicians need robust tools to assist them in this process. For the past 20 years, numerous software modules have been added to the planning systems, however none of them provide adequate collision detection (CD) and interactive system-patient setup (i.e. room level) visualization.

We proposed and presented a web-based simulation system 3DRTT [9, 10] for XRT that provides a room view of the system. The 3DRTT simulator included a generic patient model. The partial 3D model was reconstructed from patient CT data as well as from a phantom calibration tool. A wide range of analytical methods for linear accelerator-based radiation therapy (XRT) have been proposed to improve the planning process [4, 12, 13, 14]. Most methods, even though mathematically accurate, are based on hardware numeric rotational and translational values, disregarding patient-specific and detailed hardware-specific geometry. The previous research and development concerning graphical simulations of linear accelerators systems have the following essential limitations:

- Simulations use generic patient body representations [20] and inaccurate hardware 3D models, hence, potential collisions among system components and patients cannot be accurately predicted and visualized, generating patient treatment delays and additional planning costs;
- Simulations run as standalone applications and cannot be deployed over the web for immediate patient data distribution and collaboration with remote experts during treatment planning. Moreover, medical students' and trainees' access to these systems is limited.

One graphical simulation showing both the patient and hardware was developed by Hull University [1]. However, their implementation cannot be deployed freely over the web and does not have a collision detection tool. Our efforts are directed towards the development of a comprehensive yet easy to deploy web-based simulator, with the first prototype being generated in 2005. Since then, we have improved the system to provide an accurate representation of the EBRT room setup based on partial patient-specific CT data. We deployed a version of the system (3drtt.org) in 2007 and provided free registration for interested users. For the past six years we had over one thousand users on the system worldwide, most of them professionals in the medical field (Figure 2).

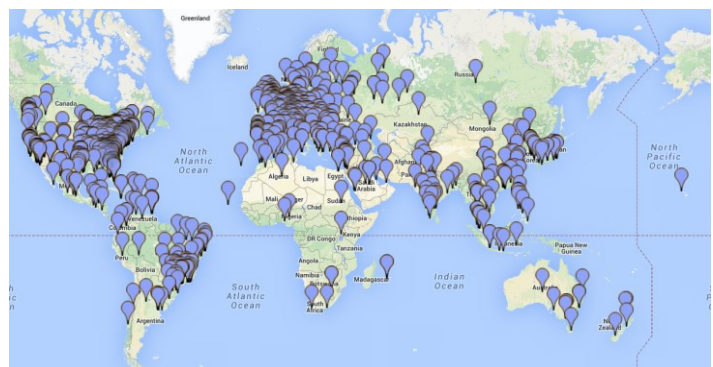
Figure 2. 1000+ 3DRTT users around the world

## 3. Patient Specific Data Acquisition

All radiation treatment plans depend on patient characteristics and need to be individualized. To improve on the targeting of the cancer and the avoidance of unnecessary irradiation to normal tissues, the planning of the radiation therapy procedure is vital.

### *3.1 Diagnostic imaging - Incomplete B-reps*

Most of the patient related data for XRT and PT is collected and stored using the DICOM RT standard [23]. Improvements in digital technology allows us to use images taken from diagnostic imaging (CT, magnetic resonance imaging (MRI), positron emission tomography (PET) scans) and incorporate them into the radiation simulation system. However only partial data is available since during diagnostic only parts of the body are scanned.

Using algorithms similar to the ones in the Visualization Toolkit [27] and a few additional optimizations techniques for rendering speed and accuracy we can extract the boundary from each CT image. We then apply the Marching Cubes algorithm [26] with a value that separates the isosurface of the patient's skin, B-Rep (Figure 3, center). The data B-rep is converted into an X3D model and embedded in the simulator (Figure 3, right).

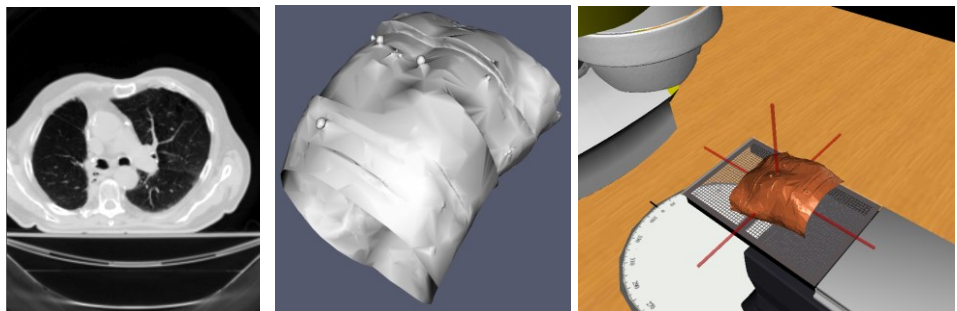

Figure 3. CT data, 3D B-Rep and X3D model in the simulator

While we have used this approach to obtain partial patient specific B-reps, full CT/PET/MRI scans are not performed on the patients due to additional radiation exposure and associated costs, hence such datasets will not provide complete patient B-reps.

### *3.2 Infrared Scanning - Complete B-reps*

We propose a cost-effective, real-time and radiation-free solution in which the patient B-Rep is obtained using commercially available infrared scanning cameras. To obtain patient specific B-reps without the additional radiation dose generated by other scanning methods, an RGB-D (red-green-blue-depth) camera is used to scan the patients. In the following sections we will present our method in more detail, describing the steps involved.

### *3.2.1 Infrared Camera Setup and Calibration*

The scanning system used in our solution is an array of up to 4 Microsoft Kinect version 2 (MKv2) infrared cameras. MKv2 has a vertical field of view of 60° and a horizontal field of view of 70° with a depth resolution of 512 horizontal by 424 vertical pixels [21]. Data acquisition occurs when the MKv2 generates a depth value (z-axis) corresponding to each pixel using the infrared sensing camera. A point cloud is the result of mapping the pixels in a depth map to corresponding x-axis and y-axis values.

Preceding the data acquisition process, calibration of the cameras is required. The calibration of the intrinsic and extrinsic camera parameters of each scanner is performed. We apply the calibration procedure using a set of checkerboard patterns as described in [8]. The patient body is primarily opaque and reflection or transparency is highly uncommon, hence the scanning process is free of reflection or transparency induced errors. However some level of infrared light interference occurs if all MKv2s run simultaneously.

### *3.2.2 Patient-Camera Configuration*

We tested several patient - scanning cameras configurations and we illustrate in Figure 4 the configuration that we currently use which includes 4 scanning systems.

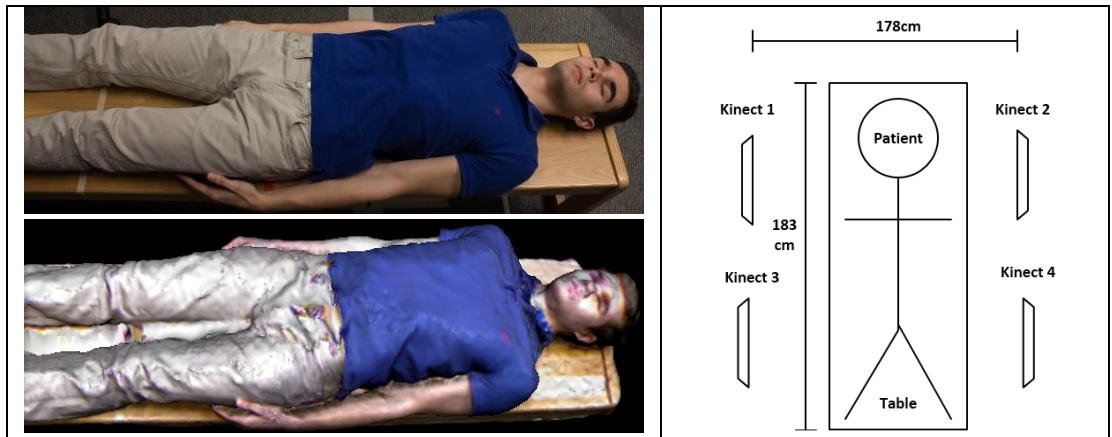

Figure 4. "Patient" on table (left-top), B-rep (left-bottom), Patient - Camera Configuration (right)

MK Developer Toolkit [22] was used for multi-static cameras to operate each MK individually (in order to reduce noise from the infrared emitters), to export the final mesh to X3D, and to take custom camera parameters. With this approach we were able to reconstruct the patient's body using 4 cameras: 2 for the upper body and 2 for the lower body (one from each side). We adjusted our algorithm to remove the data corresponding to the table by slicing the z-value of the B-rep. This scenario has the potential to generate real-time patient scans as it finalizes a scan in under 10 seconds. We have selected this scenario for two reasons:

- Hospitals with XRT and PT devices are equipped with high precision (i.e. submillimeter) translation treatment tables and the scanning cameras can be prepositioned and calibrated in advance.
- Patients undergoing the radiation therapy procedure lie on the treatment table as illustrated in Figure 4. It is preferable to scan a patient lying down as they will be doing when undergoing treatment, due to the effects of gravity on the patient's body (specifically his/her torso and abdomen). Soft tissue will usually take a different shape hence the corresponding B-rep will be different. This consideration can significantly impact the accuracy of the planning system.

### *3.2.3. Point Cloud Data Processing*

After the patient segments are acquired in the scanning phase, we use Poisson Surface Reconstruction (PSR) [16] provided by Meshlab™, to generate a new mesh. Duplicate vertices are removed during this process due to PSR's use of a voxel grid to sample the point clouds. Using a voxel grid resolution of 512 ($2^9$) we reduce the size of the data while simultaneously constructing a B-rep with minimal impact of accuracy.

A mesh optimization algorithm was employed to further reduce the size of the X3D model for web deployment. The optimization uses the Quadric Edge Collapse algorithm [18]. This algorithm is applied iteratively, with each iteration reducing the model's polygon count by 10%. Iteratively removing polygons like this results in a smoother mesh than reducing by a large number of polygons at once. In our tests, we reduced the polygon count of the mesh from 160.000 polygons to less than 16.000, which produced meshes in the range of 1.5 to 1.7 MB with texture, and .75 to .85MB without texture. Figure 5 illustrates that the boundary of the mesh remains the same after decimation. This reduced model is converted into an X3D B-rep for use in the simulator.

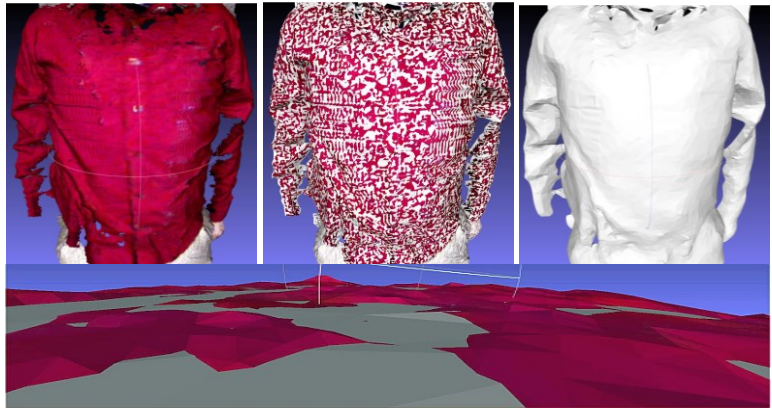

Figure 5. Original (top left), 90% decimated (top right), overlay (top middle), close up (0.5mm error) (bottom)

A discussion on the accuracy of the B-rep as it relates to the scanning system as well as it undergoes decimation and processing is presented in Section 5.

## 4. Results: Complete B-reps in the XRT/PT Simulation

An overview of the 3DRTT [9] simulator is presented followed by the presentation of the new features: (1) the Proton Therapy module and (2) the integration of patient full body scans (B-reps) in the system.

We have implemented the 3DRTT user interface using X3D [29] and an extensive library of functions written in Asynchronous JavaScript and XML (AJAX) [28] that runs in combination with Java Server Pages (JSP) [24] and provides innovative functionality and enhanced interactivity for the simulator. To mention just a few interactive components of the graphical user interface: the user can adjust the gantry, collimator and table/couch position and orientation as well as the X-Ray collimator gap (multileaf collimators can also be simulated however this feature is under development). In the first version deployed in 2007 (Figure 6 - Left) we had a generic 3D patient model while in the new version deployed in 2014 (Figure 6 – Right) we have an example of a Proton Therapy simulation with real patient data (B-rep).

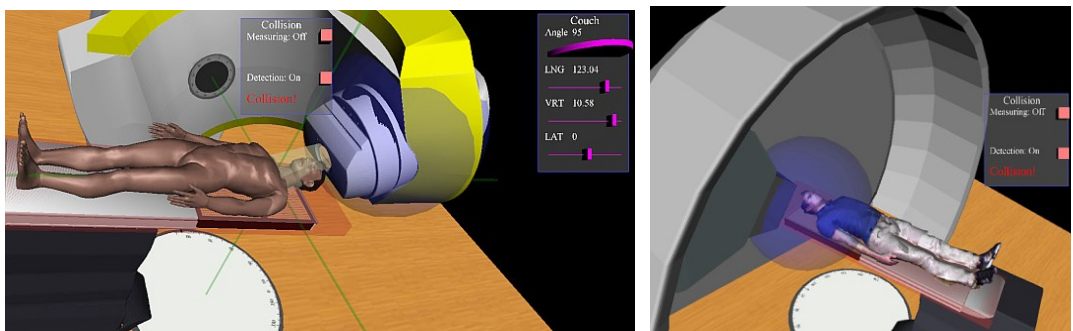

Figure 6. Collision detection: XRT with generic patient (2007) and PT with real patient (2014)

Proton Therapy devices generate similar collision problems like XRT systems. A software collision detection (CD) algorithm automatically warns the medical planning team of potential collisions among the system components and the patient, as illustrated in Figure 6. The CD system provides information about collisions in the hardware setups by highlighting the colliding objects and displaying a collision-warning message on the screen. Being able to run through a set of setups (different angles, translations, patient position etc…) the planning team can visualize in real-time potential problems and adjust the setup (angles, radiation doses etc.) before the actual radiation is delivered.

Patient specific B-reps data can be loaded into the simulator in real-time. Figure 7 illustrates different full body 3D B-reps as they are loaded into an XRT (left) and into a PT system (right) as well as the graphical user interface that allows full interactivity of the 3D scene.

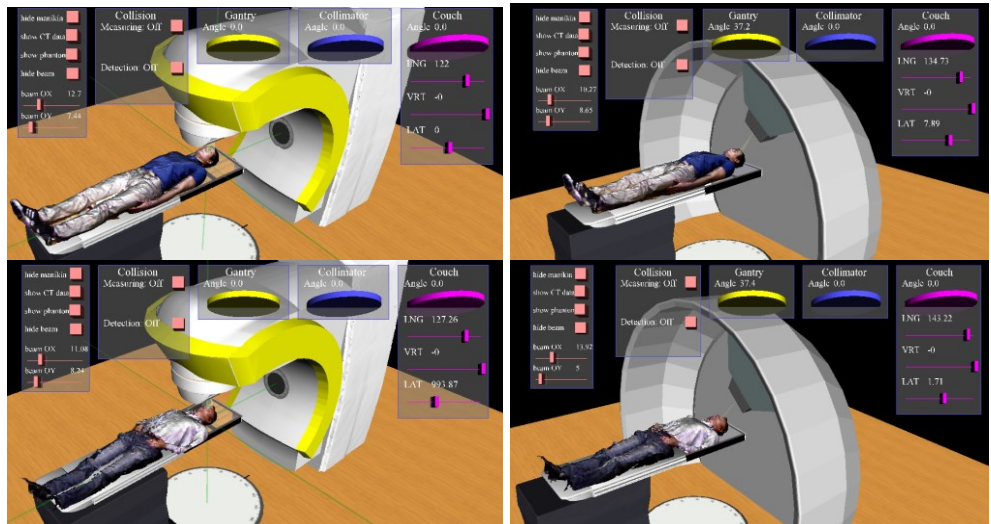
Figure 7. Graphical user interface: XRT and PT with full body B-rep

The graphical user interface allows system (collimator, gantry and table) control and navigation (translations and rotations of the view point), zooming in and out on different components of the system and the patient in order to check for potential angles, as well as comparing the relative position and orientation of components. Other functions like collimator and/or table attachment selection are also possible.

## *5. Accuracy and Precision Assessment*

Accuracy and precision assessment for radiation therapy (XRT and PT) systems is a fundamental and extremely complex task since it involves procedures like: target volume determination, organs at risk tolerance, organ movement, dose distribution and degree conformation, patient positioning and immobilization [3]. As a discussion on accuracy and precision for radiation therapy is beyond the scope of this paper, we are focusing only on the geometric accuracy of the components involved in the XRT and PT systems: the hardware 3D models and the full patient 3D models. Our accuracy assessment objectively indicates how closely the simulated geometry corresponds to the real hardware and how well the simulator matches the actual clinical setups. While the precision of the measurements is important our main focus is to provide an accurate X3D replica of the real system.

Previous research on accuracy of radiation therapy systems from the geometrical perspective investigate accuracy values based on different factors e.g. the type of tumors, the body part exposed to radiation and present accuracy requirement values in the 4 mm to 6 mm range [6]. Visual inspection of the setup before and during planning as well as during radiation delivery is very important even with the current trend of robotic systems [19]. Since surgical intervention has become a computer-mediated practice that embeds the surgeon into a complex setting of medical devices, it is no longer the patient's body but the image of the body that is the central reference for the surgeon. We prove that the 3D patient B-reps generated using our scanning system, besides being real-time, are accurate within 8- 9 mm of the real patient.

## *5.1. Hardware System Accuracy; Collision Scenarios*

The accuracy assessment objectively indicates how closely the simulated geometry corresponds to the real hardware and how well the simulator matches the actual clinical setups. Since we cannot obtain the deviation distribution for the entire geometry of the system, one solution is to run identical real and virtual (simulated) treatment scenarios and compare the clearance among the hardware components in near-collision cases. The experiments were carried out at the M.D. Anderson Cancer Center, Orlando. We considered 20 collision and near-collision scenarios, measuring (with digital calipers of submillimeter accuracy) the distances between the (potential) colliders. We reproduced these scenarios in our virtual setting and measured the same distances using a virtual measurement tool. The focus was on collimator/couch and collimator/head-fixation attachment interactions, as they are a frequent cause of collision. Figure 8 provides visual comparisons of collision scenarios between the Varian Trilogy™ and the Novalis® linear accelerators, and our corresponding simulation scenarios. The objective assessment resulted in a mean difference of 5-10 millimeters for the Varian and the Novalis® simulator, with a standard deviation of 0.5. A comprehensive description of the assessment methodology and details are available in [9].

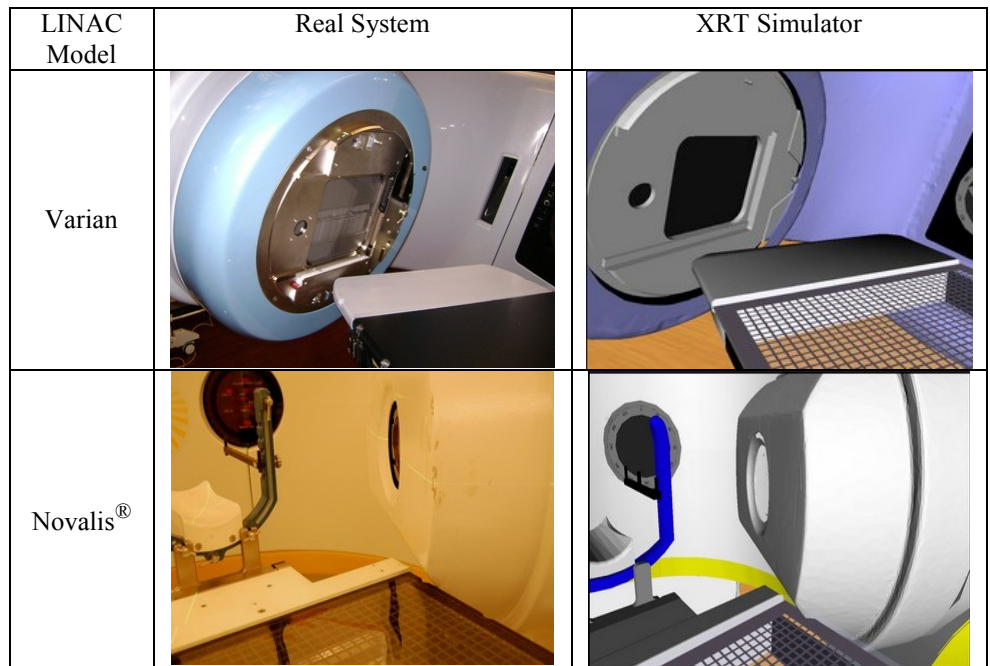

Figure 8. Illustration of real (left) and simulated (right) collision scenarios

As illustrated in Figure 8, the Varian Trilogy™ and the virtual representation have similar appearance in the collision scenarios. The B-reps for the gantry and collimator for this system were modeled based on point clouds collected with the Faro™ Technologies laser scanner [11]. For the Novalis® gantry the 3D model was built using Konica Minolta™ Vivid 3D system [17]. Radiation oncologists who collaborated with us on the 3DRTT project as well as previous research [6] consider that an overall accuracy below 1 centimeter is acceptable for collision detection purposes and will help significantly in obtaining a visual representation of the treatment equipment and patient setup, improving the confidence and quality of radiation delivery.

Ultimately, the simulator's hardware components' 3D model accuracy can be further enhanced by better polygonal model acquisition techniques (e.g., use of higher-resolution laser scanners), which will reduce geometrical errors to sub-millimeter values. A very important consideration is that high resolution 3D scanning and subsequent X3D conversion cannot be done in real-time due to scanning surface properties (the hardware equipment has shinny/reflective and transparent surfaces), however since the hardware does not change often the 3D models can be optimized offline before they are added to the simulator.

### 5.2 Patient B-Rep Accuracy

When assessing the accuracy of the patient 3D B-rep it is important to consider first the accuracy of the scanning system. Since the accuracy of the scanning system decreases with distance [2], the closer a point is to the center of the depth map the greater its accuracy. Likewise the closer a point is to the edge of a depth map the lower its accuracy. We measured the distance between the scanners and a flat level surface using a laser measuring tool with submillimeter accuracy. We collected a point cloud from each scanner and removed the points that did not correspond with the flat level surface. To keep the real-time B-rep generation manageable we are interested in FOVs which can scan the height and width of a person in 2 to 5 frames (fewer frames reduce the time needed for data acquisition). We assessed the accuracy of point clouds generated by scanning flat surfaces with the following dimensions: 1 x 0.6 meters, 0.6 x 0.6 meters, and 0.4 x 0.4 meters that would be proportional with half the size of a human body. We took scans of all the flat surfaces from 1 respectively 2 meters away.

The experiment was repeated five times and generated similar values for the Mean Absolute Error (MAE), the Root Mean Square Error (RMSE) and the Maximum Error after rounding them to the nearest significant digit (e.g. 1mm). The RMSE is below 4 mm with an average deviation between 1 to 3mm as illustrated in Figure 9.

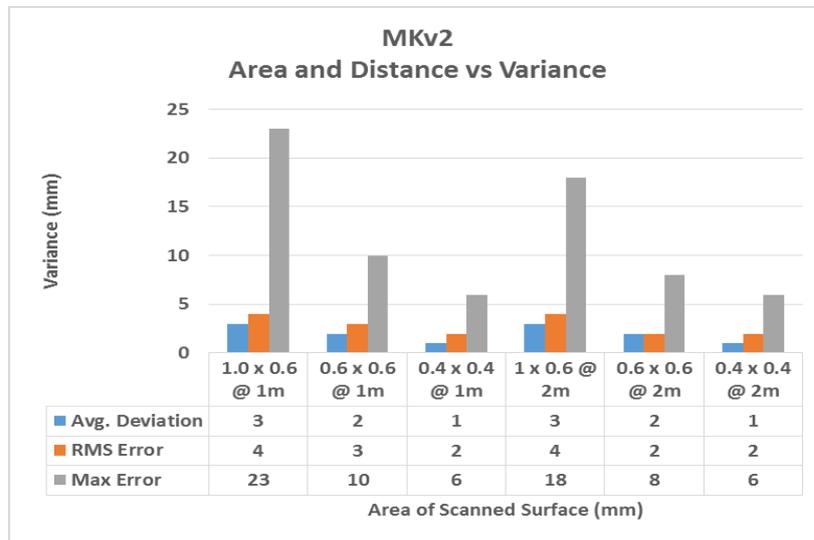

Figure 9. MKv2 Accuracy

After scanning we obtain a several point clouds that are processed using the Poisson Surface Reconstruction algorithm [16] with a RMSE of 0.001-0.002 units for the reconstruction of real world data. We consistently assessed a RMSE of 1.4 mm for our patient B-reps. After the surface is reconstructed, the Poisson algorithm creates additional geometry per its designed tendency to close holes and generate a watertight mesh. This is useful in order to close small tears in the mesh which could generate collision detection malfunction in the 3DRTT simulator, however PSR also closes data gaps that are not meant to be closed. This additional geometry must be removed. We remove inaccurate vertices by assigning a quality value based on the Hausdorff distance [5] between points on the mesh and points on the original point cloud. Then we remove values outside of 3 x RMSEs value. This allows us to filter out faces which are inconsistent with the surface of the patient or object. We found that the meshes generated using this method were within a MAE of 1.3 mm, with a maximum error of less than 3 times the RMSEs (4.2 mm). Figure 10 is a visual representation of the quality map generated, where blue represents sub-mm errors, green < 3mm and red < 5mm.

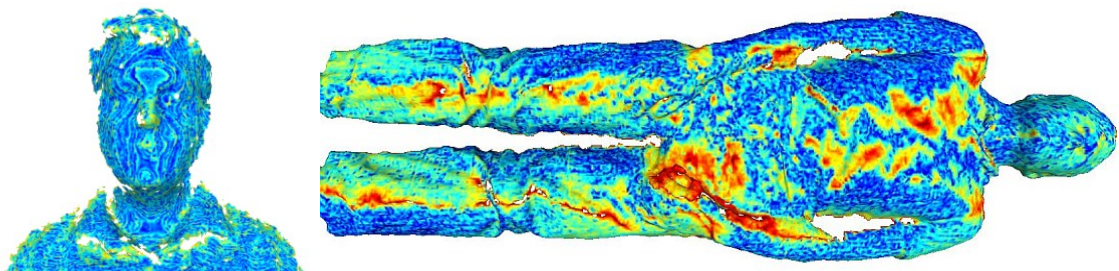

Figure 10. Quality map of a mesh generated with a MKv2. Blue <1mm, Green < 3mm, Red <5mm

These values represent excellent accuracy of the patient B-rep considering our goal of sub-centimeter accuracy on the simulation of a radiation therapy system that has a room-size volume of approximatively 27 $m^3$ that is 27x$10^6$ $cm^3$ (considering a 3x3x3m room).

While it is difficult to quantify the accuracy of a 3D B-rep obtained through scanning without a ground truth model, we can estimate an upper bound for the accuracy of the B-rep by adding errors between the different steps in the process. Between the MKv2 data acquisition and Poisson reconstruction algorithm we can generate a model with a MAE of 9.2 mm, using a 1 meter x 0.6 meter area at a distance of 1 meter from the target. However, outliers do present themselves and in ways that are difficult to detect without a ground truth model. Using the same system we can expect maximum errors to be limited, but as high as 25.2 mm. Other less convenient methods are conceivable, such as the case of a 0.4 x 0.4 meter area at 1 meter using the MKv2. Such a system would require as many as 5-10 scans depending on the length and width of the patient, but could still be automated, quick, and offer a maximum error as low as 10.2 mm.

## 6. Conclusions

We have presented a cost-effective solution for generating in real-time full patient 3D models (B-reps) as well as a new module that simulates a Proton Therapy device. The benefits of using the online XRT and PT simulator are educational and clinical. For the education component, the targeted trainee groups include physics residents, dosimetry trainees, and radiation therapy technicians. The online web-based system will allow for:

- Familiarization with the translational and rotational motion limits of various hardware components, including the couch, gantry, and collimator, and with their angle conventions for both X-Ray and Proton Therapy devices.
- Validation of patient setups, planning of deliverability and checks for possible collision scenarios and beam-couch intersections.
- Education of patients about their treatment delivery technique, to help reduce pre-treatment anxiety.

The fact that the simulator runs in a Web browser with minimal software installation allows medical personnel to easily use and adapt to the simulator graphical user interface. Moreover the system proposed enables data distribution and collaboration with remote experts during treatment planning with minimal cost and effort. This reduces costs associated with travel and data distribution and improves overall hospital efficiency related to radiation therapy procedures.